\documentstyle[prl,twocolumn,aps,epsfig,epsf]{revtex}

\begin{document}
\wideabs{
\draft
\title{\bf{\LARGE{Dynamics of an electron in finite and infinite one 
dimensional systems in presence of electric field}}} 
\author{B. C. Gupta and P. A. Sreeram }
\address{Institute of physics, Bhubaneswar - 751 005, India}
\maketitle
\begin{abstract}
We study,numerically, the dynamical behavior of an electron in a two site 
nonlinear system driven by dc and ac electric field separately. We also study,
numerically, the effect of electric field on single static impurity and 
antidimeric dynamical impurity in an infinite 1D chain to find the strength 
of the impurities. Analytical arguments for this system have also been given. 
\end{abstract} 

\vspace{.1in}

\pacs{PACS numbers : 71.10. + x; 73.20.Dx; 03.65. - w.}
\narrowtext
}
\section{Introduction}
The dynamics of an electron in presence of electric field, in a two level 
system as well as in a linear chain has been a 
subject of recent interest. \cite{miloni,gross1,cren}
It is well known that for a double-well heterostructure in the absence of 
driving forces the electron can visit either well due to quantum tunneling. 
However, if the double well structure is biased, i.e. driven by a dc field the
electron in an eigenstate will be naturally localized in one of the wells
because of the superposition of the coherent tunneling between two wells.
\cite{wang1}
When the double well system is exposed to a pure ac field the electron gets
frozen in the initially populated well, provided the ratio of the field
strength to field frequency becomes a root of Bessel's function of order
zero.\cite{gross2} In presence of both ac and dc field,\cite{wang2} the 
electron gets trapped in the initially populated site when
the ratio of the ac field strength to the frequency is such that the 
$n^{th}$ order Bessel function becomes zero and the dc field strength 
becomes the $n^{th}$ multiple of the frequency of the ac field, provided 
the energy difference between the levels is much lesser than the frequency of 
the ac field i.e., $J_n(x_n) = 0, x_n = E/\omega$, $E_0 = n \omega ~~ 
n=0,1,2.....$ and $\Delta/\omega << 1$, where $\Delta$ is the energy difference 
between the two levels. Here we consider the dynamics of an electron in a 
two site nonlinear system in presence of ac and dc electric fields separately.

On the other hand a lot of studies have been done on the dynamics of a 
charged particle subjected to a periodic potential in presence of dc, 
ac as well as in the presence of both dc and ac electric fields. It is 
well known that, in presence of dc field, a charged particle executes 
oscillatory motion in the reciprocal lattice. \cite{bloch} This, in turn, 
confines the particle spatially. \cite{wann}

In presence of ac field, the particle is generally delocalized except for
the cases when the ratio of the field magnitude to the field frequency
is a root of the ordinary Bessel function of order zero.  \cite{dunlap}
In presence of both ac and dc field, the localization induced by 
the dc field can be suppressed by the ac field when the stark frequency 
becomes equal to an integral multiple of the frequency of the ac field.
\cite{naz1}

Recently, Nazareno et. al \cite{naz2} had studied the dynamics of an 
electron in
a 1D chain in the presence of a single static impurity subjected to
a dc electric field. They had shown that when the impurity potential
( $\epsilon_0$ ) is such that it coincides with the on-site energy due the 
field ( $E$ ) on a particular site $n$, i.e.,
$eEan = \epsilon_0$, where $e$ is the charge of the particle and 
$a$ is the lattice parameter, the packet oscillates resonantly between 
the impurity site and the particular site $n$.
Therefore, when the electric field $E = \epsilon_0/ea$, the charged particle
oscillates resonantly between the impurity site and it's neighboring site.
However, we find that this condition is valid only if the particle
is strictly confined to the impurity site and its neighboring site. On the
other hand, in a 1D chain there is a probability for the particle to tunnel 
to other sites. Because of this tunneling probability the resonance condition 
gets modified.
  
We also study the dynamics of an electron in a 1D perfect chain in presence
of two nonlinear impurities \cite {ken1,ken2} which are next to each other, 
subjected to
dc electric field. We suggest a possible means of measuring the strength
of the nonlinear impurity.  

The organization of the paper is as follows. In Sec. II, we deal with
a two site nonlinear system, in presence of electric fields. In Sec.
III, we consider a one dimensional infinite chain with a static impurity
and the dynamics of an electron in presence of a dc electric field. In 
Sec. IV, we consider a 1D chain with two consecutive nonlinear impurity
sites and the dynamics of an electron in presence of a dc electric field.
Finally, in Sec. V we give a summary of our investigations.

\section{Two Site System}

The time evolution of the electron in a two site nonlinear system in presence
of both ac and dc fields is governed by the equations given by,
\begin{eqnarray}
i\frac{dC_1}{dt} & = & V C_2 + (\frac{-E_0}{2}-\frac{E}{2}{\rm 
{\rm cos}}(\omega t)
+\chi_1
|C_1|^2)C_1 \nonumber \\
i\frac{dC_2}{dt} & = & V C_1 + (\frac{E_0}{2}+\frac{E}{2}{\rm cos}(\omega t)
+\chi_2
|C_2|^2)C_2 .
\label{difeq2site}
\end{eqnarray}
Here $C_1$ and $C_2$ are the probability amplitudes of the particle to stay at
site 1 and 2 respectively. $V$ is the hopping amplitude of the particle 
between the two sites, $E_0$ is the amplitude of the dc field, $E$ is
the amplitude of the ac field and $\omega$ is the frequency of the ac field.
The kind of nonlinearity we consider here arises from the interaction of the
electron with the vibration of the local oscillators in the system \cite{ken1}.
$\chi_1$ and $\chi_2$ are the nonlinear parameters determining the strength
of the interaction of the particle with the local oscillators at site 1 and
2 respectively. 

In the absence of both dc and ac electric fields (i.e. $E_0=E=0$),
the two site system is completely isolated from any external perturbations
and the dynamics of the particle is governed completely by the strength
of the nonlinearity at the two sites and the hopping matrix element
between the two sites.

In the special case when the strengths of nonlinearity at the two sites
are equal (i.e. $\chi_1=\chi_2=\chi$), there exists a critical value of $\chi$ 
( $\chi_{cr} = 4 V $), below which the particle remains delocalized between 
the two sites and above which the particle gets trapped at the initially 
populated site. \cite{ken1} Thus, there is a abrupt transition from a 
completely
delocalized state to a completely localized one across the critical
value of the nonlinear parameter defined by $\chi_{cr}$.

The other interesting situation arises when $E_0 = E = 0$ and 
$\chi_1 = -\chi_2 = \chi$. This kind of nonlinearity appears when both 
the local oscillators oscillate in opposite phase and interact with the 
particle with similar strength. In this case, the time evolution of the 
probability of the particle at the initially populated site is given by,
\begin{equation}
P_1(t) = \frac{\chi^2}{\chi^2+4V^2} + (1-\frac{\chi^2}{\chi^2+4V^2})
{\rm cos}(\omega t) .
\label{prob1}
\end{equation}
In this limit, the transition is a continuous one, where, as the nonlinearity 
increases, the particle continuously gets trapped at the initially populated 
site. This is obvious from Eq.\ (\ref{prob1})  where $P_1 (t) \rightarrow 1$ 
as $\chi \rightarrow \infty$.

We can, on the other hand, put $\chi_1=\chi_2=0$ and $E=0$. This is
the case when nonlinearity is absent and a pure d.c. field is applied to
the two site problem. In this case, the two site problem effectively
reduces to a two level problem, where the degeneracy of the two
 levels ( due to equal site energies) is lifted by the externally applied
dc electric field. It has been shown that, in this situation, the
particle remains localized at the initially populated site provided the
hopping matrix element is small compared to the strength of the electric
field (i.e. $V/E_0 << 1$).

In the absence of nonlinearity and dc electric field ($\chi_1=\chi_2=0,
E_0=0$) and in the presence of an ac electric field ( given by 
$E{\rm cos}(\omega t)$), the particle remains delocalized between the two sites
except when the ratio of the amplitude of the a.c. filed ($E$) to its
frequency ($\omega$) is such that the zeroth order Bessel function
$J_0(E/\omega)$ is identically equal to zero. Here too, the condition
for localization is satisfied only when $V/E <<1$.\cite{rag}

The question naturally arises as to what happens to the dynamics of the
electron in the nonlinear two site system in presence of field. Here, we 
consider the case where the  nonlinear parameters are of equal strength 
but are opposite in sign. As mentioned earlier, this kind of situation 
arises when the local oscillators oscillate in opposite phase but interact 
with the particle with same strength. Therefore, we consider $\chi_1 = \chi$ 
and $\chi_2 = -\chi$.
In this situation, the site energies of both the sites vary with time. 
Now, we consider the effect of a dc and ac electric field on the dynamics of
an electron in the nonlinear two site system, separately. 

In presence of a pure dc electric field of strength $E_0$, the time evolution
of the probability of the electron in the system is governed by the equations,
\begin{eqnarray}
\frac{dP}{dt} && = 2VR \nonumber \\
\frac{dQ}{dt} && = -(E_0-\chi)R \nonumber \\
\frac{dR}{dt} && = -2VP + (E_0-\chi)Q 
\label {PQReqns1}
\end{eqnarray}
where $P = \rho_{11}-\rho_{22}$, $Q = \rho_{12}+\rho_{21}$, $R = 
i(\rho_{12}-\rho_{21})$ and $\rho_{ij}=C_iC_j^\star$, i,j=1,2. 
On solving Eqs.\ (\ref{PQReqns1}), the probability difference of the particle 
between the two sites is given by,
\begin{equation}
\rho_{11}-\rho_{22} = \frac{(E_0-\chi)^2}{\Omega^2}+ \frac{4V^2}{\Omega^2}
{\rm cos}(\Omega t)
\label{2sitedens}
\end{equation}     
where $\Omega = \sqrt{4V^2+(E_0-\chi)^2}$. For the sake of discussion, we 
take $\chi$ and $E_0$ to be positive. Thus Eq.\ (\ref{2sitedens}) explicitly 
shows that the  particle becomes fully delocalized only when $\chi = E_0$. 
The field direction can be reversed by replacing $ E_0$ by $-E_0$. In this
case however, there is no delocalization for any field strength, since 
$\chi$ is still positive.

We now consider the case of a pure ac electric field applied to the same 
nonlinear two site system. The dynamics of the particle in this system is 
governed by the following equations.
\begin{eqnarray}
\frac{dP}{dt} & = & 2VR  \nonumber \\
\frac{dQ}{dt} & = & -(E{\rm cos}(\omega t) -\chi)R \nonumber \\
\frac{dR}{dt} & = & -2VP + (E{\rm cos}(\omega t) -\chi)Q.
\label{PQReqns2}
\end{eqnarray}
where $P, Q$ and  $R$ are defined earlier. We cannot decouple these equations
and hence solve them numerically, using fourth order Runge-Kutta method. 
We have checked the probability conservation of the particle at each time 
step. We find that, 
as long as $V/E << 1$, the particle becomes fully delocalized only when 
$\chi = n\omega$ where $n$ is a positive integer and $E/\omega$ is such that
the $n^{th}$ order Bessel function becomes nonzero (i.e. $J_n(E/\omega) 
\neq 0$). On the other hand, the particle gets fully trapped at the initially 
populated site only when $\chi=n\omega$ and $E/\omega$ is such that the 
$n^{th}$ order Bessel function is identically equal to zero (i.e. 
$J_n(E/\omega) = 0$). 

     \begin{center} 
      \epsfig{file=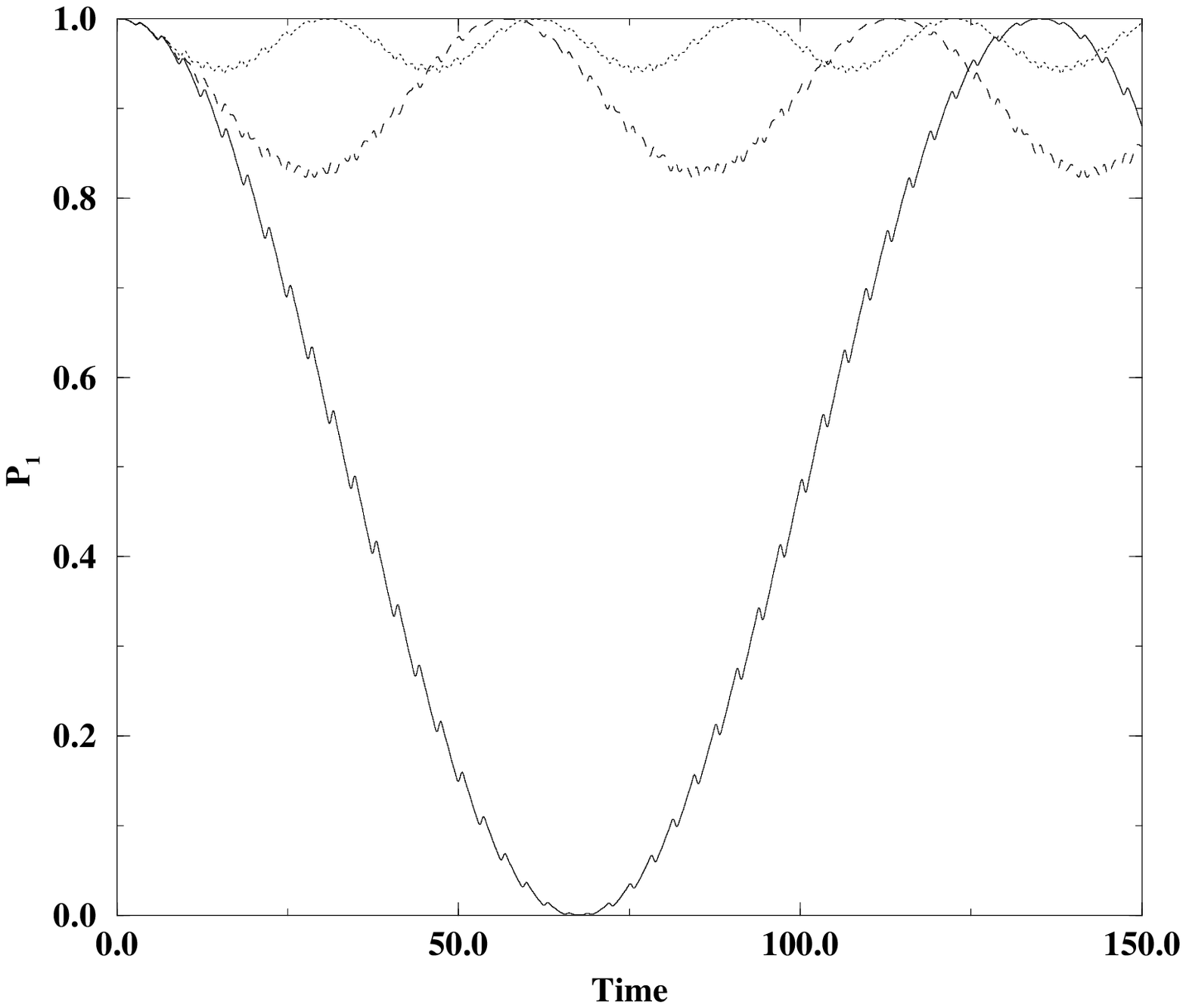,width=9cm,height=7cm}
      \vspace{-1.0cm}
      \begin{figure}[p]
      \caption{The probability of the electron at site 1 ($P_1$) as a function 
              of time, where, $V=0.8, \omega=2, E/\omega=1.9155$ such that 
              $J_1(E/\omega) \neq 0$. The dotted
              curve, solid curve and the dashed curve are for $\chi$=1.8, 2.0
              and 2.1 respectively.
      \label{probki}} 
      \vspace{0.5cm}
      \end{figure}
      
      \epsfig{file=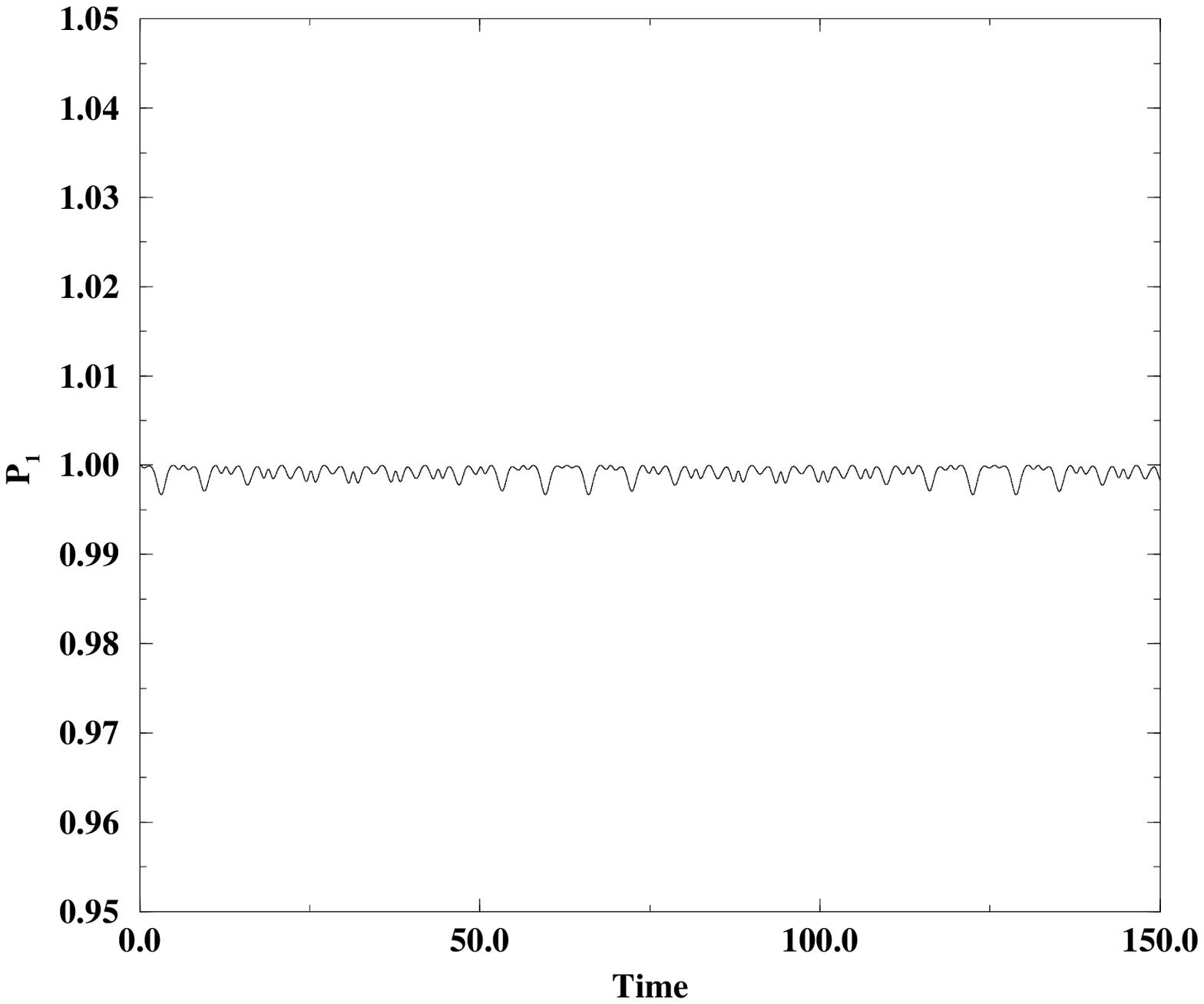,width=9cm,height=7cm}
      \vspace{-1.0cm}
      \begin{figure}[p]
      \caption{The probability of the electron at site 1 ($P_1$) as a function 
              of time, where, $V=0.8, \omega=\chi=1, E/\omega= 3.831$, such 
              that $J_1(E/\omega)=0$. 
      \label{probki1}} 
      \vspace{0.5cm}
      \end{figure}
      \end{center}
To show this, we have plotted $P_1= \frac{1+P}{2}$ 
( probability of the particle to stay at the site 1)
as a function of time in Figs.\ (\ref{probki}) and (\ref{probki1}). 
In Fig.\ (\ref{probki}), we have taken $V=0.8,
\omega=2$ and $E/\omega = 1.9155$ such that $J_1(E/\omega) \neq 0$. The 
dotted curve corresponds to $\chi = 1.8 < \omega$, the solid
curve is for $\chi = 2.0 = \omega$ and the dashed curve for $\chi = 2.1
> \omega$. The dotted and dashed curves show that, as long as the nonlinearity 
strength ($\chi$) is not an integral multiple of the frequency ($\omega$) of 
the field, the electron remains mostly localized at the initially populated 
site. As soon as the nonlinearity parameter matches with the $n^{th}$ multiple
of the frequency of the ac field (solid curve), the electron gets fully 
delocalized. On the other hand, in Fig.\ (\ref{probki1}), we have taken 
$V=0.8$, $\omega= \chi=1$
and $E=3.831$ such that $J_1(E/\omega)=0$. The figure clearly shows that
the electron gets completely frozen at the initially populated site. 
Thus, a two site system in presence of this kind of nonlinearity and ac field
gives the same kind of dynamical behavior of the electron as is the case
of a two level system subjected to dc and ac fields simultaneously. Here,
the nonlinearity plays the role of a dc field in a two level system.

The complete delocalization of the electron in this system can be understood 
by the following physical reasoning. Because of
the presence of nonlinearity, the site energies of the two sites become
different. Thus, the nonlinearity maps the two site system to a two
level system. Since site energy at both the sites depends on the probability
of the particle to stay at the respective sites, they fluctuate with time.
However, the energy difference between the sites, $\Delta = \chi(\rho_{11}
+\rho_{22}) = \chi$ which is constant. This is because the total probability
of the particle in the system is conserved. Therefore, when the particle is in 
one of the sites it can absorb an amount of energy $\chi = n \omega$, where
$n$ is an integer, from the ac field and jump to the other site and thus, 
full delocalization is obtained for this condition.  

\section{1D Chain with Single Static Impurity}

In this section, we consider a one dimensional infinite perfect chain with a 
single static impurity at the middle (0$^{th}$) site. Our interest here is
to find the dynamics of an electron initially populated at the 
impurity site in presence of a dc electric field. The governing
equation for the electron dynamics in presence of the dc electric field
is given by a set of differential equations, as,
\begin{eqnarray}
i\frac{dC_n}{dt} = \epsilon_n \delta_{n,0} C_n + eaE_0nC_n 
+ V(C_{n+1} + C_{n-1}), \nonumber \\
~~~n=-N,-N+1,......,N-1,N. \label{infinty}
\end{eqnarray}
$C_n$ is the probability amplitude of the particle to stay at the
$n^{th}$ site, $\epsilon_n$ is the site energy at the $n^{th}$ site,
$e$ is the electronic charge, $a$ is the lattice spacing, $E_0$ is the
strength of the dc electric field and $V$ is the hopping integral between
neighboring sites. $N$ is chosen such that the particle does not feel
the boundaries, which in turn ensures the infiniteness of the chain.
Without any loss of generality, we choose $e=1$,$a=1$ and normalize all
energies with respect to the hopping matrix element $V$. The delta
function implies that the site energy of all sites except the zeroth site 
have been taken to be zero, in absence of the dc electric field. To
know the time evolution of the electron probability at site $n$ we need
to solve Eqs.\ (\ref{infinty}). Since, it is not possible to solve them
analytically, we use fourth order Runge Kutta method to solve them
numerically. Here also, conservation of the probability of the particle has 
been checked at each time step. 
     \vspace{-1.0cm}
     \begin{center} 
      \epsfig{file=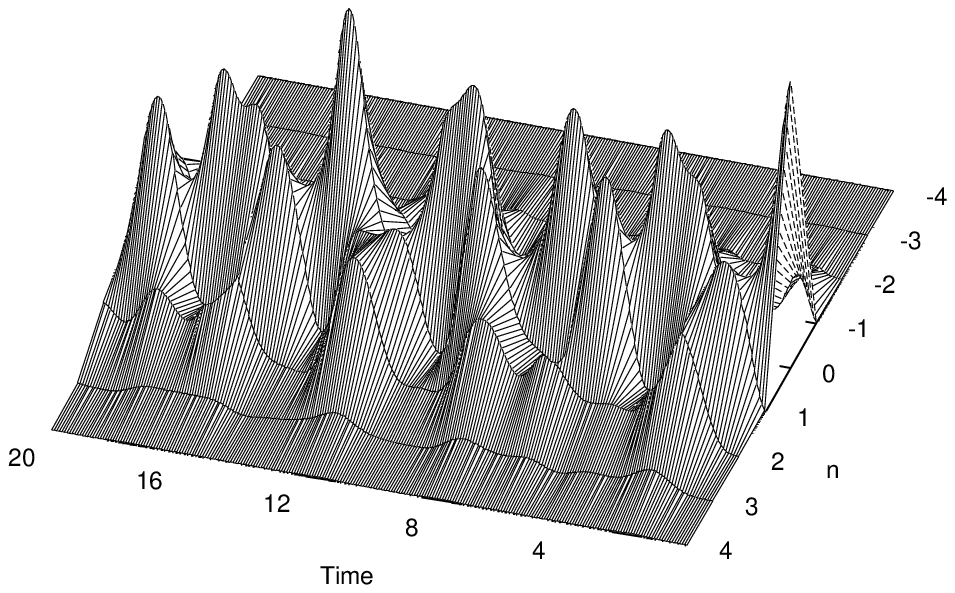,width=9cm,height=7cm}
      \vspace{-1.0cm}
      \begin{figure}[p]
      \caption{Time evolution of the probability of the electron in
               a 1D chain with a static impurity at the $0^{th}$ 
               site. $n$ represents the site index. The z-axis
               represents $|C_n|^2$. Here
               $\epsilon_0$ (strength of the impurity)= $E_0$ (strength of the 
               electric field) = 2.  
      \label{k2e2}} 
      \vspace{0.5cm}
      \end{figure}
      \end{center}
      \vspace{-1.0cm}
      \begin{center}	
      \epsfig{file=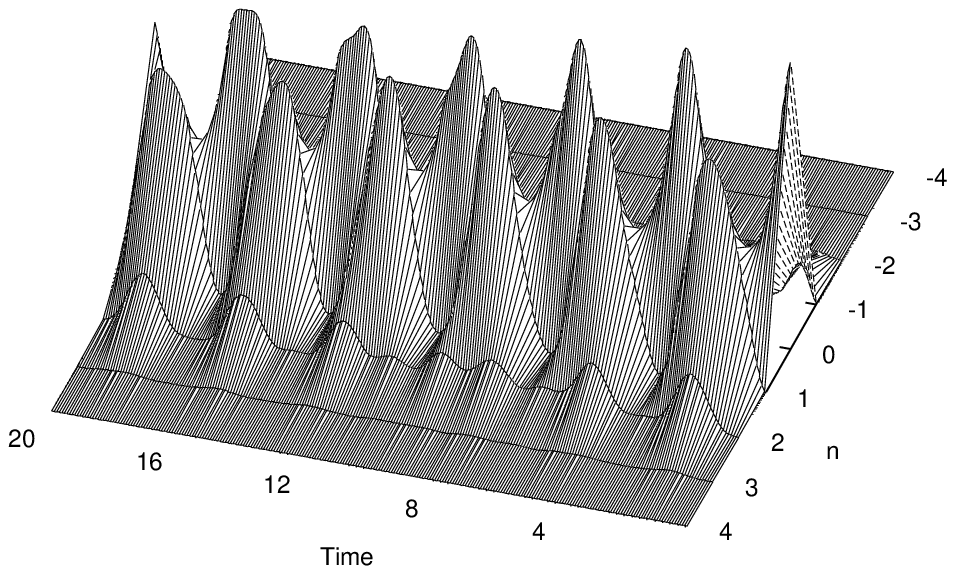,width=9cm,height=7cm}
      \vspace{-1.0cm}
      \begin{figure}[p]
      \caption{Same as Fig.\ (\ref{k2e2}) except $E_0$ = 2.67.  
      \label{k2e267}} 
      \vspace{0.5cm}
      \end{figure}
      \end{center}
For a fixed value of $\epsilon_0$ there is a particular 
value of the electric field strength ($E_0^{res}$) at which the 
electron oscillates resonantly between the impurity site and one of its 
nearest neighbor sites. We observe that, when $\epsilon_0$ is small, 
then the electric field strength required for the resonance condition is 
larger than $\epsilon_0$. 
We also find that, for smaller value of $\epsilon_0$,
the departure of this electric field strength ($E_0^{res}$) from $\epsilon_0$ 
is larger and it decreases
as $\epsilon_0$ increases. Thus, for very large values of $\epsilon_0$,
the resonance condition reduces to $E_0^{res}=\epsilon_0$. We plot
the time evolution of the probability of the electron in the system
for $E_0=\epsilon_0=2$ in Fig.\ (\ref{k2e2}) and the same for $E_0=2.67$,
$\epsilon_0=2$ in Fig.\ (\ref{k2e267}). 
It is obvious from Fig.\ (\ref{k2e2}) that for $E_0=\epsilon_0=2$, the 
electron propagates through the system and gets confined among many 
sites. On the other hand, Fig.\ (\ref{k2e267}) shows that the particle 
oscillates
mainly between the impurity site and one of its nearest neighbors.
Therefore, for $\epsilon_0=2$, $E_0=2.67$ is the electric field required
for resonance between the impurity site and its neighbor. 
     \begin{center} 
      \epsfig{file=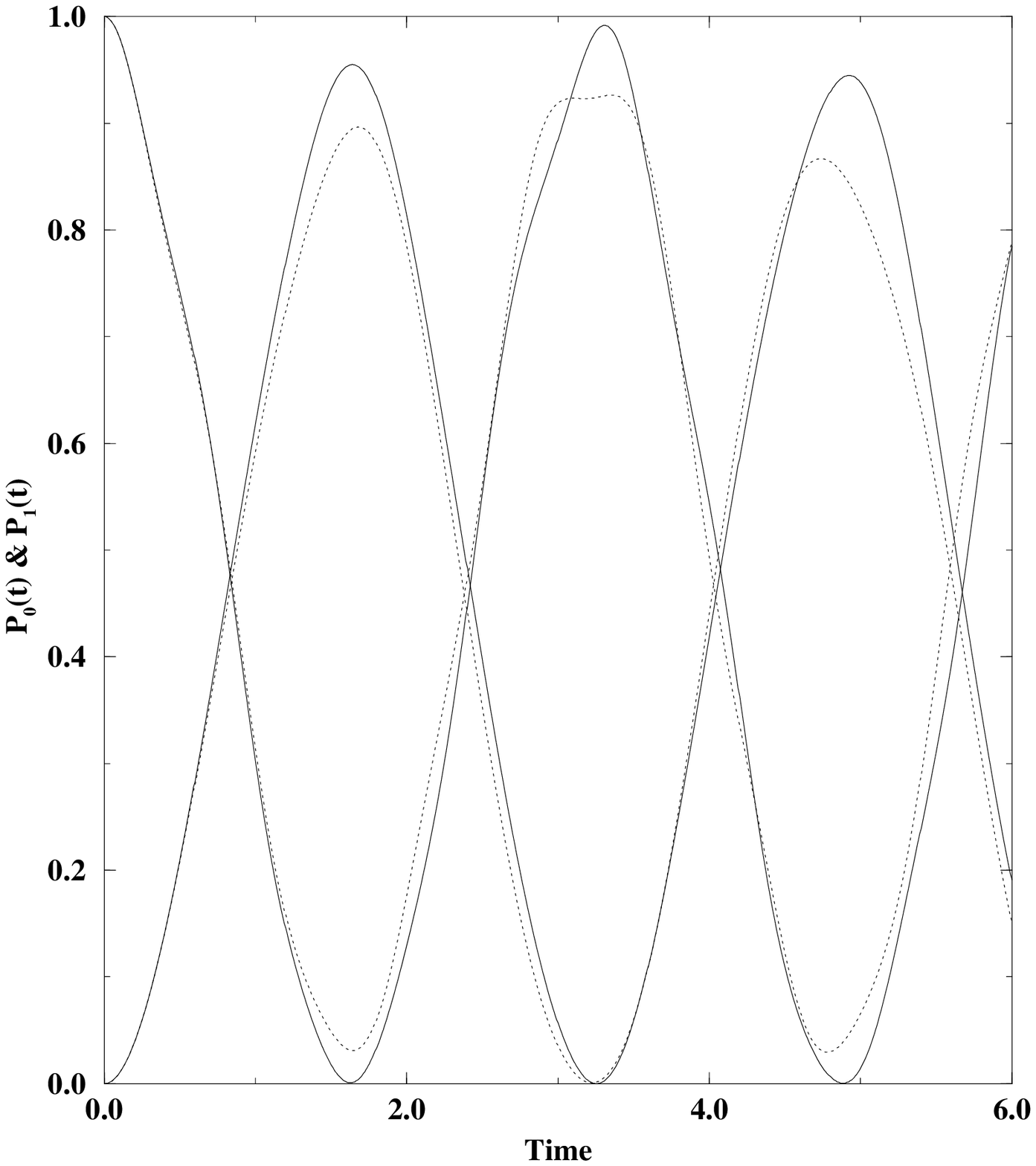,width=9cm,height=7cm}
      \vspace{-1.0cm}
      \begin{figure}[p]
      \caption{ Probability of the electron at site 0 ($P_0$) and at
                site 1 ($P_1$) as function of time. The solid curves
                are for $\epsilon_0=4.0$, $E_0=4.37$ and the dotted curves
                are for $\epsilon_0=4.0$, $E_0=4.0$.
      \label{compk4}} 
      \vspace{0.5cm}
      \end{figure}
      \end{center}
In Fig.\ (\ref{compk4}) we
have also plotted the probability of the electron at the impurity site and
its nearest neighbor site as a function of time for $\epsilon_0=E_0=4.0$
(dashed curves) and $\epsilon_0=4.0$,$E_0=4.37$ (solid curves).
Here too, it is clear, that for $\epsilon_0=4.0$ and $E_0=4.37$ the particle
oscillates more strongly between the impurity site and its nearest neighbor
than for $\epsilon_0=E_0=4.0$. Therefore, for $\epsilon_0=4.0$ one needs
the field strength, $E_0=4.37$, to get the resonant motion of the particle
between the impurity site and its neighbor. 
In order to understand this
deviation from the normally expected resonance condition (i.e. $\epsilon_0=
E_0$) we find the resonance condition for general impurity strength as a 
function of electric field strength analytically. For this, we need to 
confine the particle between the impurity site (site 0) and its nearest 
neighbor site (site 1). To do this, we renormalize the hopping elements 
between the sites 0 and -1, and between sites 1 and 2 respectively. 
This in 
turn renormalizes the site energies at sites 0 and 1. The effective hopping 
between sites 0 and -1 and that between sites 1 and 2 thereby decreases. The 
effective energies at site 0 and 1 are given by
\begin{eqnarray}
\epsilon_0^{eff} & = & \epsilon_0+V^2/E_0 
\nonumber \\
\epsilon_1^{eff} & = & E_0-V^2/(2E_0)
\label{effens}
\end{eqnarray}
By putting $\epsilon_0^{eff} = \epsilon_1^{eff}$ we get the condition for 
resonance given by
\begin{equation}
E_0^{res} = \frac{\epsilon_0 + \sqrt{\epsilon_0^2 + 6}}{2}
\label{reson1}
\end{equation} 
This condition is in contrast to the condition for resonance given
by Nazareno et.al. \cite{naz2} Eq.\ (\ref{reson1}) explicitly shows that the 
resonance condition reduces to $\epsilon_0=E_0$ only when $\epsilon_0 >> 
\sqrt{6}$.   Thus, to get a resonant motion of the particle between
the sites 0 and 1, a higher value of electric field is required as 
compared to the impurity strength. We have plotted $E_0^{res}-\epsilon_0$
as a function of $\epsilon_0$ obtained from Eq.\ (\ref{reson1}) (solid
curve) along 
with the corresponding numerical plot ( dotted line) in 
Fig.\ (\ref{static}). 
     \begin{center} 
      \epsfig{file=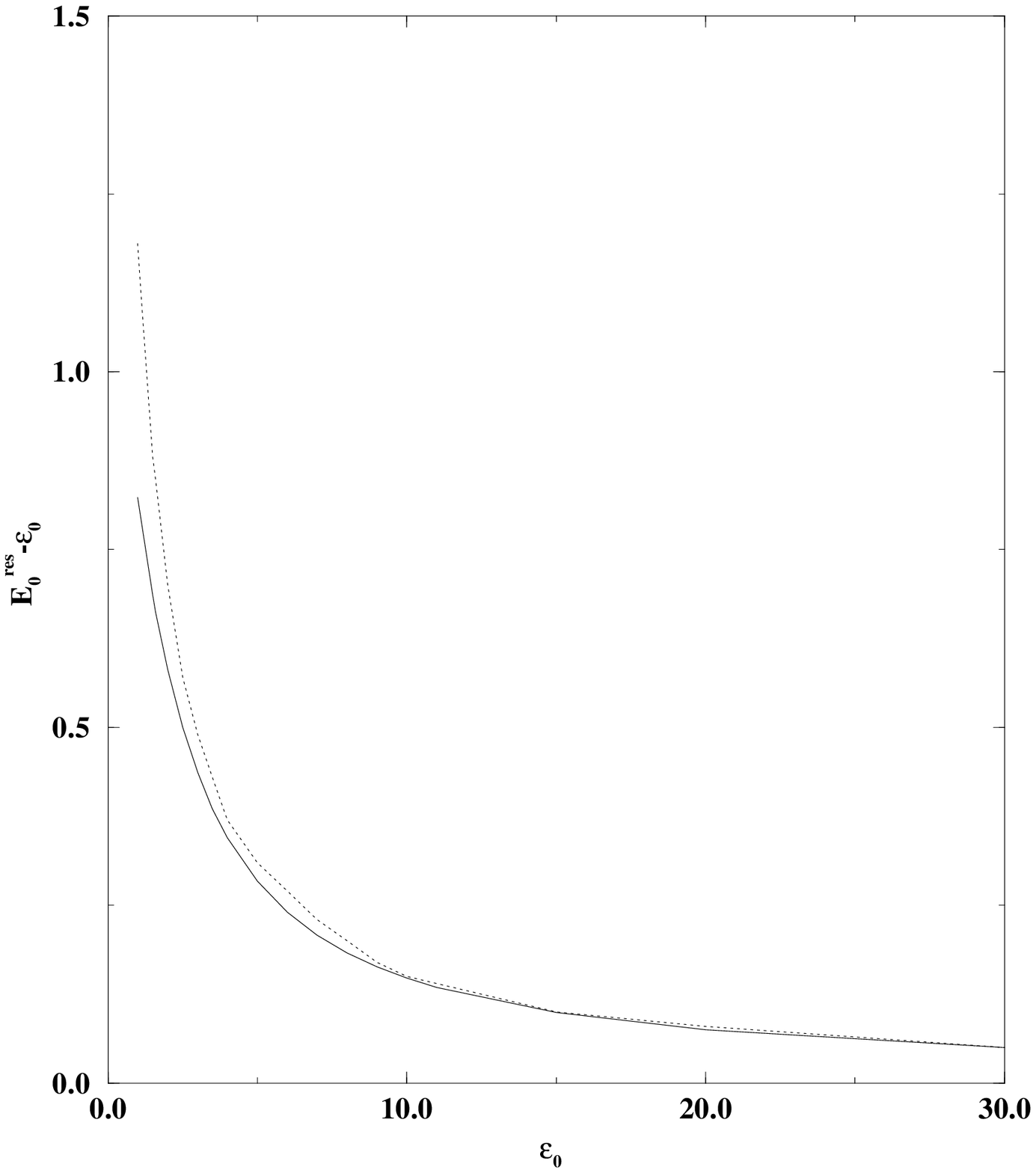,width=9cm,height=7cm}
      \vspace{-1.0cm}
      \begin{figure}[p]
      \caption{$E_0^{res}-\epsilon_0$ as a function of $\epsilon_0$. The
               dotted curve and the solid curve represent the numerical
               and analytical estimates, respectively. 
      \label{static}} 
      \vspace{0.5cm}
      \end{figure}
      \end{center}
We see a good agreement 
between the numerical and analytical estimates. One can estimate the higher
order corrections to Eq.\ (\ref{reson1}) by further renormalizing the hopping 
matrix elements. However, the contribution from the higher order corrections
is negligibly small. We therefore, restrict ourselves only upto the first 
order correction. Upto the first order,
the strength of the static impurity can be evaluated using Eq.\
(\ref{reson1}), once there is resonant oscillation between the impurity
site and its nearest neighbor, which is characterized by the emission
of a photon for the particular value of the dc electric field ($E_0^{res}$),
as remarked by Nazareno et.al. \cite{naz2}

\section{Nonlinear impurities in a chain}

In this section we consider a system of 1 dimensional chain with two nonlinear
impurities at two consecutive sites. The nonlinearity arises due to
the interaction of the particle with lattice vibration. We consider a
special case where the particle interacts with the local oscillator only
when it is at site 0 or site 1. We denote the middle site as the 
$0^{th}$ site. The interactions at both the sites ( site 0
and 1) are equal in strength but opposite in nature. This may arise
only when the oscillators at sites 0 and 1 oscillate in opposite phase.
Assuming the presence of this kind of nonlinearity, we ask whether it is 
possible to find the strength of the interaction, by applying an external 
electric field, as was done for the single static impurity case in the 
previous section. Here, we give an answer to this question. 

We start by applying  a pure dc electric field to the system and see the 
time evolution of the electron populated initially at the $0^{th}$ site. The 
dynamics of the electron is governed by the equations given by,
\begin{eqnarray}
i\frac{dC_0}{dt} & = & \chi |C_0|^2 C_0 +V(C_1+C_{-1})\nonumber \\
i\frac{dC_1}{dt} & = & (-\chi |C_1|^2 + E_0) C_1 +V(C_0+C_{2}) \nonumber \\
i\frac{dC_n}{dt} & = & n E_0 C_n +V(C_{n+1}+C_{n-1}),~~~n \ne 0,1.
\label{diffeqlinch}
\end{eqnarray}
We solve these equations numerically by using fourth order Runge
Kutta method. We check the conservation of probability here too.
We observe that, as $E_0$ increases, the particle becomes 
more and more confined between sites 0 and 1. The amplitude of oscillation
between sites 0 and 1 increases and then at some $E_0$ ( say, $E_0^r$)
it becomes maximum. If $E_0$ is increased further, the particle gets trapped 
at the initially populated site. So, at $E_0=E_0^r$ the particle oscillates 
between sites 0 and 1 resonantly. Here also, as is the case with a single 
static impurity in an infinite chain, for any strength of nonlinearity,
$\chi$ the electric field needed to get the resonance condition is higher
than $\chi$. This situation can also be detected by the radiation emitted by
the electron while oscillating resonantly between sites 0 and 1. 

We now find the resonance condition analytically. For this we have to
reduce the system into an effectively two site system. In other words we 
need to
confine the motion of the particle between sites 0 and 1. Since,
at resonance condition, the time average probability of the particle
at sites 0 and 1 are approximately 1/2, we therefore replace the
site energies arising from nonlinearity at site 0 and 1 by $\frac{\chi}{2}$ 
and $\frac{-\chi}{2}$ respectively. We then renormalize the hopping matrix 
element between sites 0 and its left neighbor and that between site 1 and
its right neighbor ( the sites are numbered in increasing order from left 
to right, with the middle site being the zeroth site). As a consequence of 
this renormalization, the site energies at sites 0 and 1 get renormalized 
and are given by,
\begin{eqnarray}
\epsilon_0^{eff} & = & \frac{\chi}{2}+\frac{V^2}{E_0} \nonumber \\
\epsilon_1^{eff} & = & \frac{-\chi}{2}+E_0-\frac{V^2}{2E_0}
\label{nonlinchain1}
\end{eqnarray}
Now equating these effective site energies we get the the condition for the 
particle to oscillate resonantly between site 0 and 1. This condition is 
given by
\begin{equation}
\chi=E_0^r-\frac{3V^2}{2E_0^r}.
\label{nonlinchain2}
\end{equation}
     \begin{center} 
      \epsfig{file=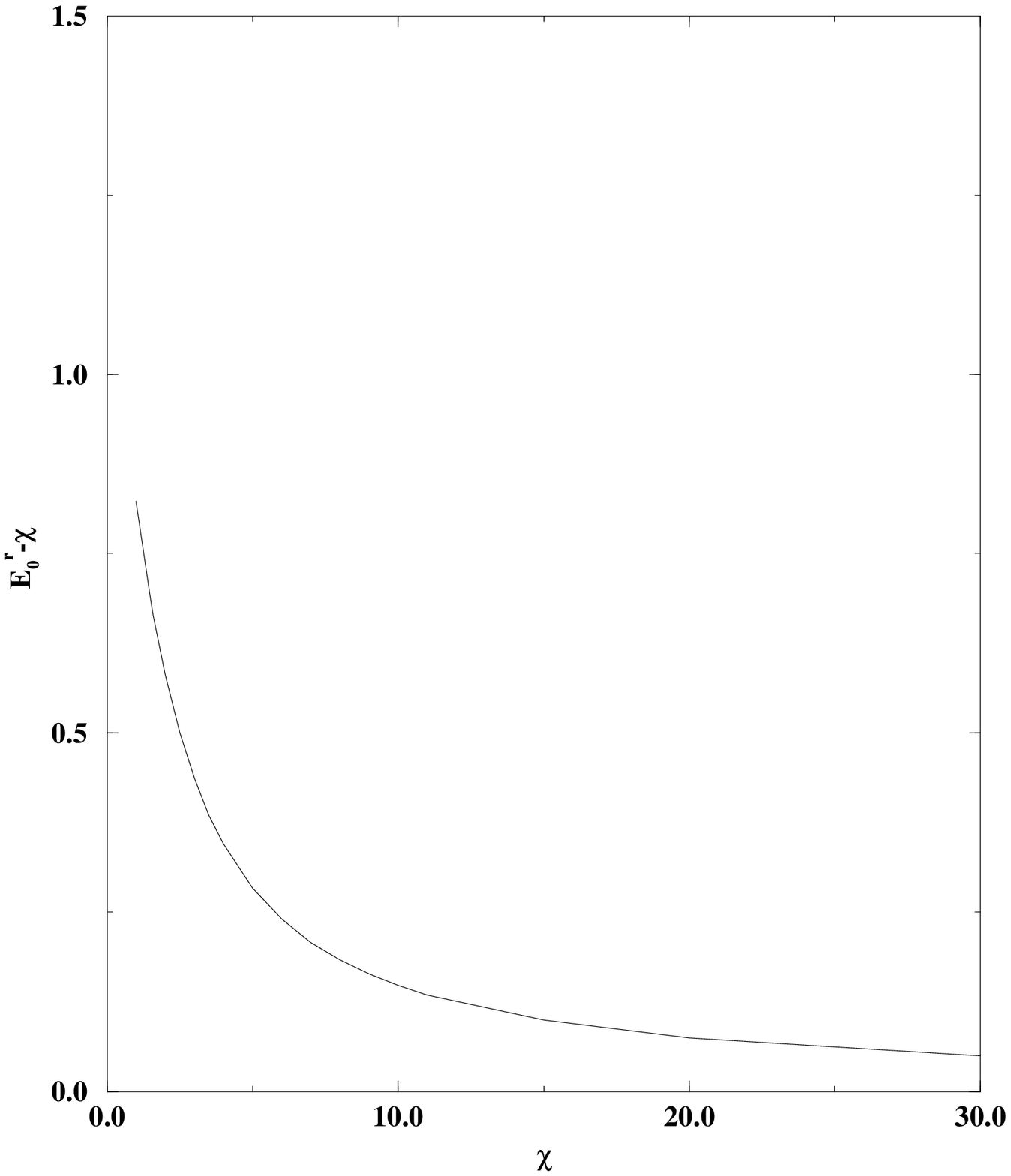,width=9cm,height=7cm}
      \vspace{-1.0cm}
      \begin{figure}[p]
      \caption{$E_0^{r}-\chi$ as a function of $\chi$. 
      \label{nonlin}} 
      \vspace{0.5cm}
      \end{figure}
      \vspace{-1.0cm}
      \epsfig{file=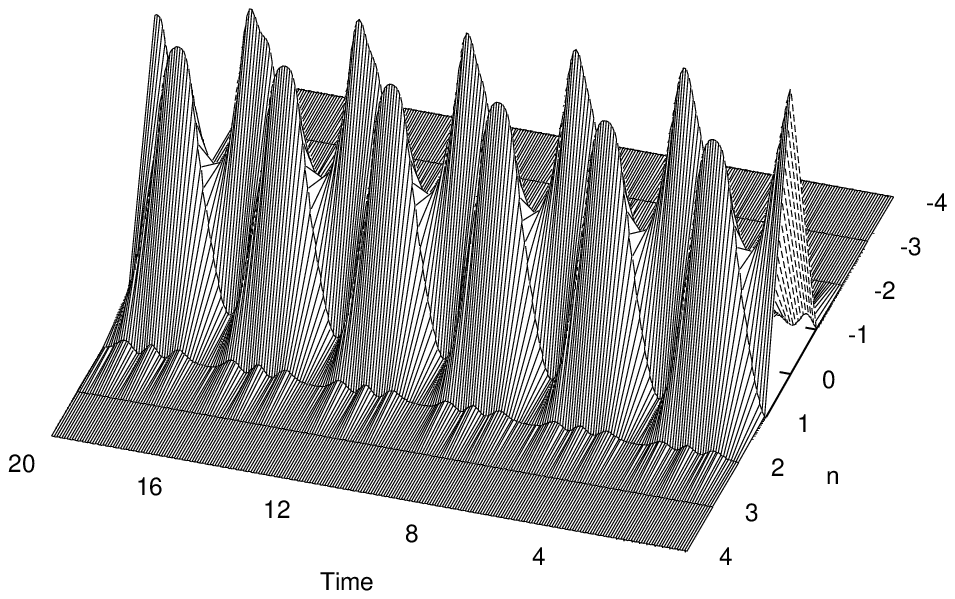,width=9cm,height=7cm}
      \vspace{-1.0cm}
      \begin{figure}[p]
      \caption{Time evolution of the probability of the electron in
               a 1D chain with nonlinear impurity $\chi$ 
               at site 0  and -$\chi$ at site 1.
               $n$ represents the site index. The z-axis
               represents $|C_n|^2$. Here
               $\chi$ = 4.0 and $E_0$ = 4.34.
      \label{k4e434}} 
      \vspace{0.5cm}
      \end{figure}
      \end{center}
We plot $E_0^r-\chi$ as a function of $\chi$ obtained 
analytically in Fig.\ (\ref{nonlin}). Now we take a point from the curve for 
resonance
condition (Fig.\ ({\ref{nonlin})) which corresponds to $\chi = 4.0$ and 
$E_0 = 4.34$.
For these values of $\chi$ and $E_0$ we see the the time evolution of the
electron probability in the system. This is shown in Fig.\ (\ref{k4e434}). 
We notice that 
the electron mostly oscillates between sites 0 and 1 resonantly. This also
confirms our analytical estimate for resonance condition.
Thus tuning the electric field one can 
obtain the resonance condition and hence the strength of the nonlinear impurity
can be estimated using Eq.\ (\ref{nonlinchain2}) provided that special kind
of nonlinearity is present in the system.
\section{Conclusion}
We have studied the dynamics of an electron in a two site nonlinear system
in presence of externally applied dc and ac electric fields, separately. 
We have shown the localization-delocalization conditions in both the
cases. We have also studied the dynamics of an electron in a infinite 1D
chain having a single static impurity in presence of a dc electric field.
We have given an analytic expression for the condition when the particle 
oscillates resonantly between the impurity site and its nearest neighbor.
This, in turn, gives a possible method of measuring the impurity strength.
A similar analysis has been performed for an antidimeric dynamical impurity
present in an otherwise perfect chain. 
\section{Acknowledgment}
We thank Dr. S. Sil for useful discussions and suggestions.
 
\end{document}